# Fundamental parameters of some B-type stars using NOAO Indo-U.S. Library

A. Shokry[1,2], M. I. Nouh[1], S. M. Saad[1,2] and I. Helmy[1]

abdo_nouh@hotmail.com

[1]Astronomy Department, National Research Institute of Astronomy and Geophysics (NRIAG), 11421 Helwan, Cairo, Egypt.

[2]Kottamia Center of Scientific Excellence in Astronomy and Space Science (KCScE, STDF, ASRT), Cairo, Egypt

**Abstract**: In the present article, we present a spectroscopic analysis of 83 field B-type stars from NOAO Indo-U.S. Library. We calculated the fundamental parameters e.g effective temperatures, surface gravities, and rotational velocities using Barbier-Chalonge-Divan (BCD) method and line blanketed LTE/NLTE model atmospheres. The projected rotational velocity was estimated by fitting the Mg II4481 Å line profile with theoretical lines calculated from LTE/NLTE models. The evolutionary masses for the program stars are estimated using the evolutionary models. In most of the cases, the present study gives fair agreement with earlier investigations and is even more accurate in some cases.

1.  Introduction

A long-standing and significant concern in astrophysics is calculating the physical properties of stars. Benchmarks against calculated models of stellar structure and evolution are the masses, radii, and effective temperatures of stars. Due to significant advances in astronomical instruments and observational techniques, spectral analysis of stellar spectra has been the subject of widespread interest in the past few decades. The fundamental stellar parameters computed from libraries lead to a wide and more specific understanding of the population synthesis and spectra of galaxies.

Examples of these libraries are: STELIB Stellar Library (Le Borgne et al., 2003) is a homogeneous visible library of stellar spectra covers the spectral range (3200 to 9500 Å), including stars of all spectral types, luminosity classes, and metallicity. The MILES stellar library (Sánchez-Blázquez et al. 2006 and Falc´on-Barroso et al., 2011) consists of ~1000 stars covering a wide variety of atmospheric parameters. The library covers the spectral range 3525 Å - 7500Å, secured by Isaac Newton Telescope (INT) at FWHM =2.50 Å.

The B-type stars are rather hot, massive stars, those stars with visible spectra dominated by neutral He lines and hydrogen lines of the Balmer series, *while the metallic lines are weak or absent.* The effective temperature of B-type stars is in the range 10000 -30000 K° for B9- B0 respectively. They are believed to be young and moving relatively rapidly through the first stages of evolution.

Most of our knowledge about stellar atmospheres of B-type stars comes from basic physical parameters (such as effective temperature $T_{eff}$ and surface gravity log g) of individual stars. Accurate $T_{eff}$ is needed for locating stars on the HR diagram and for abundance determinations, Sokolov (1995). It can be derived by indirect methods based on the comparison of observed quantities (such as the color index, the flux distribution, and the line profiles) with the corresponding computed ones.

In the present paper, we mainly deal with the determination of the fundamental parameters e.g. $T_{eff}$, log g, using, and spectral types of some field B star based on the Indo-US library. The calculations by Huang & Gies (2008) depended on the LTE model atmospheres without line blanketing. As demonstrated by many investigations, line blanketing affects the calculations of effective temperatures. This is due to the presence of lines in the star's spectrum affecting both the energy balance in the outer layers of the star and the distribution of energy. The latter effect can be divided into two parts: (a) a true redistribution of continuum spectral energy induced by the shift in the energy balance in the star, and (b) an evident redistribution of spectral energy induced by several overlapping lines by the 'obscuration' of the continuum (Athay, 1972).

We applied the method of the Barbier-Chalonge-Divan (BCD) to calculate the effective temperatures and surface gravities of 83 field B- type stars, which were previously investigated by Huang & Gies (2008). The method of BCD has many advantages, where the parameters (D, λ1) are obtained from direct measurement on the stellar continuum energy distribution, also the extinction of stars could be neglected as demonstrated by Zhořec et al (2009). The rotational velocities are estimated by fitting the Mg II 4481 Å line profile with theoretical ones. Section 2 is devoted to discussing the properties of the observed and synthetic spectra, in sec. 3 we briefly described the methods of analysis, the results are given in section 4, and section 5 concluded our analysis.

## 2. The Observed and Synthetic Spectra

### 2.1. Observed Spectra

The observed spectra are selected from the NOAO Indo-U.S Archive of Coudé Feed Stellar Spectra (Valdes et al. 2004). The spectra of 1273 stars were carried out using the 0.9 m Coudé Feed telescope at Kitt Peak National Observatory, at a low resolution of~ 1.2 A° FWHM (Valdes et al. 2004), in the spectral range $\lambda\lambda$ 3460 - 9464 A°. Nearly 140 B-star spectra are found in this archive, after checking them, we found only 83 spectra are suitable to apply the BCD method. The rest of the stars were excluded because their spectra have bad calibration, or they were out of the wavelength range of BCD. Figure 1 displays a sample of normalized spectra of the three program stars.

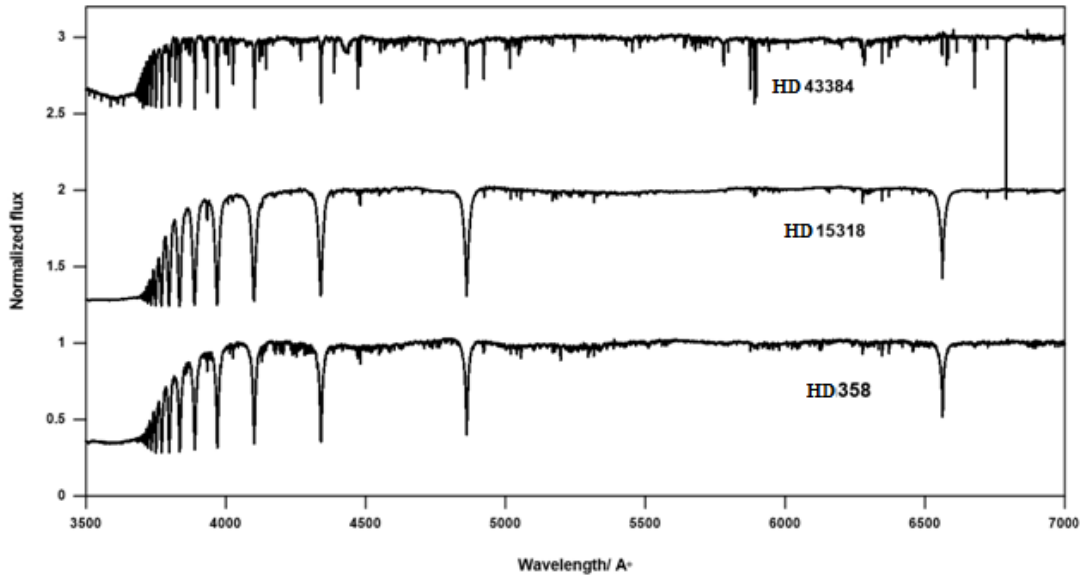

Figure 1: Normalized spectra for three of the NAOA Indo-Us B-stars.

### 2.2. Synthetic Spectra

For the purpose of the spectral analysis, we used LTE/NLTE grid suitable for the effective temperature range of B-type stars. We adopted ATLAS9 grids (Kurucz, 1992) as input models for LTE calculations, assuming a solar metallicity, a microturbulent velocity of 2 km/s, and a mixing length to scale height ratio of 1.25. The effective temperatures span the range in the model grid of 250 K for stars cooler than 10000 K, rising to 25000 K for hotter stars. The models have surface gravities $1 \leq \log g \leq 5$. LTE spectra (for the range λλ 1500 A° - 8000 A°) are synthesized using the SPECTRUM code written by Gray (1992, 1993). SPECTRUM inputs the columns for mass depth points, temperatures, and total pressure, then calculates using a system of seven nonlinear equilibrium equations at each stage in the atmosphere.

For NLTE synthetic spectra calculations we adopted the BSTAR2006 model atmosphere grid calculated by Lanz and Hubeny (2007). The grid consists of sixteen effective temperature values ($15000 \leq T_{eff} \leq 30000$ K) with 1000 K steps, 13 surface gravities ($1.75 \leq \log g \leq 4.75$) with 0.25 dex steps, six chemical compositions, and a microturbulent velocity of 2 km/s are considered. In Figure 2 we plotted the normalized synthetic spectra of some program stars labeled with effective temperatures and surface gravities, one can notice the magnesium line (Mg II 4481 Å) which could be considered as a valuable indicator of the rotational velocities.

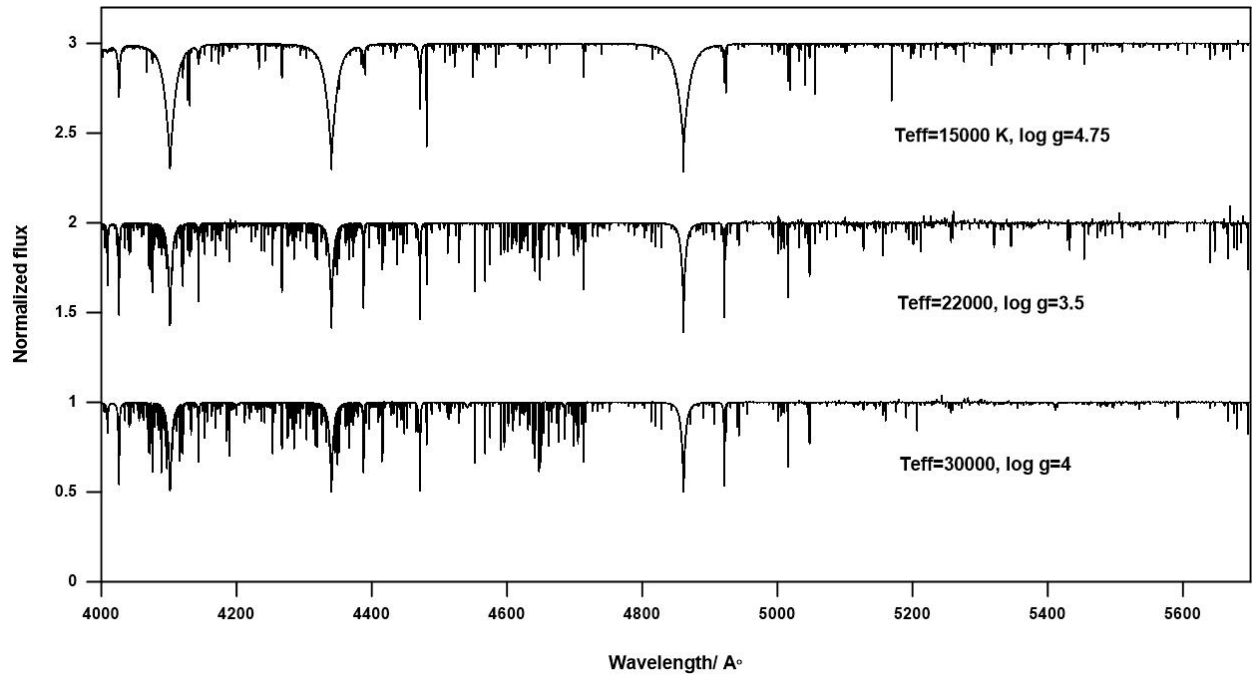

Figure2: Normalized synthetic spectra of three of the B-type stars, labeled with effective temperature and surface gravities.

## 3. Method of Analysis

We used the BCD method to determine effective temperatures and surface gravities of the program stars. The BCD method was originated by Barbier & Chalonge (1941) and Chalonge& Divan (1952). It is based on measuring of two parameters (D, λ1) of the Balmer discontinuity, where D is the height of the Balmer discontinuity measured by extrapolating the Balmer and Paschen continua to a wavelength of 3700 A°, while λ1 is the mean spectral position of the Balmer discontinuity, at (λ -3700). Figure (3) shows the measurable parameters of BCD method D and λ1, where the red and blue lines represent the interpolation of the upper (Paschen) and lower (Balmer) continuum spectrum, respectively.

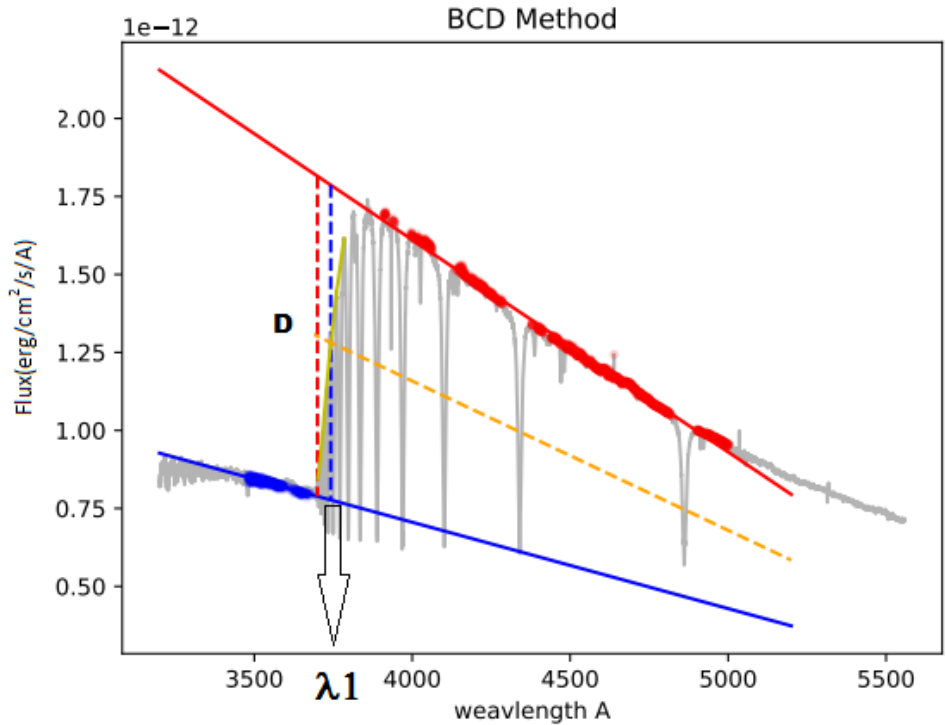

Figure 3: Graphical explanation of D and λ1 parameters of the BCD method.

To apply the BCD method, we measured D and λ1 for each spectrum by using the BCD method, more details about the method are given in the work of Shokry et al.(2018). To determine the $T_{eff}$ and log g, then we fitted these measurements to the theoretical grid of the Bruce model (Townsend et al., 2004) to obtain the best parameters. The errors in $T_{eff}$ and log g are calculated using the uncertainty of the estimated D and λ1 (due to the uncertainties related to the continuum placement). The parameters D and λ1 are found by interpolating the points representing the maxima and minima of the continuum. Moreover, the Teff and log(g) were obtained relying on the estimated parameters. The determination of the maxima and minima is a tricky task. Consequently, we change the maximum and minimum levels by using a fixed step, in which the two levels are incremented or decremented in a reverse direction with the same value. This process is repeated several times, such that the step is a multiple of the iteration number. Furthermore, each iteration leads into two different Teff and log(g) values. Eventually, the error is the standard deviation of these values.

Using the code ROTIN3 (I. Hubeny, http://tlusty.oca.eu/Synspec49/synspec-frames-rotin.html), we calculated a grid of rotated spectra, then by applying the $\chi^2$ criterion we estimated the best rotational velocity. The projected rotational velocities (v sin i) are derived by fitting the Mg II 4481 Å observed line profile to the synthetic one calculated from LTE/NLTE models (Mg II 4481 Å recommended to be free from the pressure broadening, (Gray, 1976)). We used the LTE model for effective temperatures lower than 15000 K (Kurucz, 1992), while the NLTE models are used for stars higher than 15000 K (Lanz and Hubeny, 2007).

## 4. Results

### 4.1. Fundamental parameters of the studied stars

The results obtained for the program stars are listed in Table 1; where columns 1, 2, 3, and 4 represent the name of the star, spectral type, the height of the Balmer discontinuity (D), and the mean spectral position (λ1); columns 5, 6, 7 and 8 list the effective temperatures, surface gravities, and rotational velocities, respectively.

Based on the effective temperatures and surface gravities of the program stars computed by the BCD method, we selected the suitable theoretical synthetic spectra from the LTE/NLTE grid and generate a rotated grid using the ROTIN3 routine. By applying $\chi^2$ we estimated the best fitting rotated model, Figure (4) displays the best match of the observed MgII 4481 Å and the synthetic one. We obtained a good fit for most of the program stars, except for the line core of the stars HD 87344 and HD 185423 and the line wing of the star HD 171301.

**Table 1: Stellar parameters of the studied B stars from NOAO archive.**

| HD | Sp. Type | D | $\lambda_1$ | $T_{eff}$ (K) | | log g (dex) | | $V_{rot}$ sin i (km/s) | | $M/M_\odot$ |
|---|---|---|---|---|---|---|---|---|---|---|
| | | | | Present | HG | Present | HG | Present | HG | |
| 4727 | B7V | 0.38 | 45.65 | 12065 ± 438 | ---- | 3.85 ± 0.073 | --- | 60±2.45 | --- | 3.5± 0.12 |
| 10362 | B5III | 0.23 | 32.2 | 15528 ± 248 | 13211±133 | 3.40 ± 0.005 | 3.144±0.024 | --- | 61±11 | 7.0±0.11 |
| 12303 | A0III | 0.54 | 36.68 | 9756 ± 31 | 11491±81 | 3.27 ± 0.005 | 3.195±0.023 | 60±0.21 | 77±12 | 4.0± 0.01 |
| 17081 | B8V | 0.41 | 59.14 | 11837 ± 31 | 12769±89 | 4.03 ± 0 | 3.689±0.023 | — | 5±20 | 3.5± 0.01 |
| 25940 | B5IV | 0.24 | 38.92 | 15363 ± 260 | 17746±552 | 3.61 ± 0 | 3.898±0.060 | 200±3.38 | 166±10 | 6.0± 0.10 |
| 27295 | B9V | 0.49 | 65.91 | 10452 ± 127 | 11334±113 | 4.05 ± 0 | 3.972±0.036 | 20±0.24 | 33±15 | 2.5± 0.03 |
| 34797 | B9V | 0.46 | 65.91 | 10998 ± 199 | ---- | 4.05 ± 0.004 | --- | — | --- | 3.0± 0 |
| 35497 | A0IV | 0.53 | 56.89 | 9867 | 13129±98 | 3.73 | 3.537±0.023 | — | 60±5 | 3.0± 0 |
| 38899 | B9V | 0.47 | 61.39 | 10670 ± 86 | 10272±40 | 4.03 ± 0.005 | 3.781±0.018 | — | 39±4 | 3.0± 0.03 |
| 41692 | B8V | 0.41 | 52.39 | 11579 ± 74 | 13669±144 | 3.97 ± 0.025 | 3.26±0.020 | 10±0.08 | 37±12 | 3.5± 0.03 |
| 58343 | B5IV | 0.24 | 36.68 | 15461 ± 284 | 15025±317 | 3.53 ± 0.028 | 3.428±0.045 | — | 35±10 | 6.0± 0.12 |
| 74280 | B5IV | 0.24 | 43.4 | 15261 ± 209 | 18630±411 | 3.76 ± 0.040 | 3.933±0.050 | --- | 101±5 | 5.0± 0.07 |
| 75333 | B9V | 0.48 | 59.14 | 10605 ± 52 | 12105±121 | 3.97 ± 0.005 | 3.775±0.036 | 20±0.10 | 49±16 | 3.0± 0.05 |
| 79158 | A0IV | 0.54 | 59.14 | 9722 ± 129 | 12718±228 | 3.73 ± 0.065 | 3.554±0.056 | 30±0.65 | 57±12 | 3.0± 0 |
| 79469 | B9V | 0.47 | 63.65 | 10779 ± 2.5 | 10190±39 | 4.04± 0.015 | 3.92±0.022 | 110±0.02 | 93±7 | 3.0± 0.04 |
| 87344 | A0V | 0.53 | 56.89 | 9921 ± 11 | 10689±64 | 3.75 ± 0.050 | 3.526±0.026 | — | 32±9 | 3.0± 0.05 |
| 100889 | B9V | 0.47 | 61.39 | 10708 ± 363 | 10422±38 | 4.03 ± 0.073 | 3.649±0.018 | 210±8 | 235±10 | 3.0± 0.10 |
| 109387 | B5V | 0.27 | 41.16 | 14177 ± 57 | ---- | 3.70 ± 0.035 | --- | 190±1.9 | --- | 5.0± 0.17 |
| 120315 | B4V | 0.24 | 47.89 | 15374 ± 541 | 15689±128 | 3.90 ± 0.135 | 4.004±0.022 | 170±8.4 | 144±5 | 5.0± 0.17 |
| 129956 | B9V | 0.46 | 59.14 | 10888 ± 147 | 10333±51 | 4.01 ± 0.025 | 3.731±0.023 | — | 87±7 | 3± 0 |
| 147394 | B7V | 0.4 | 61.39 | 12193 ± 240 | 14166±149 | 4.04 | 3.806±0.026 | 30±6 | 0±15 | 3.5± 0.08 |
| 149630 | B9IV | 0.47 | 54.64 | 10694 ± 129 | 10600±34 | 3.89 ± 0.055 | 3.598±0.017 | 260±4.8 | 276±15 | 3.0± 0 |
| 150100 | B9V | 0.46 | 63.65 | 10882 ± 111 | 10441±42 | 4.04 | 4.015±0.017 | 10±0.1 | 79±12 | 3.0± 0.05 |
| 150117 | B9V | 0.46 | 59.14 | 10815 ± 92 | 10594±37 | 4.00 ± 0.044 | 3.67±0.017 | 200±2.7 | 203±10 | 3.0± 0.06 |
| 155763 | B6IV | 0.29 | 34.44 | 13535 ± 606 | 12833±86 | 3.48 ± 0.069 | 3.543±0.020 | 10±0.5 | 47±4 | 5.0± 0.22 |
| 157741 | B9V | 0.46 | 59.14 | 10810 ± 98 | 10569±43 | 4.00 ± 0.024 | 3.639±0.020 | 260±2.8 | 287±13 | 3.0± 0.04 |
| 158148 | B7V | 0.36 | 45.65 | 12381 ± 912 | 14210±99 | 3.85 ± 0.047 | 3.733±0.017 | — | 247±6 | 4.0± 0 |
| 160762 | B3V | 0.18 | 50.14 | 18392 | 15961±155 | 3.94 | 3.613±0.025 | 10±0 | 5±2 | 6.5± 0 |
| 161056 | B2IV | 0.14 | 38.92 | 20705 ± 921 | 20441±327 | 3.59 ± 0.108 | 3.433±0.039 | 280±15 | 287±8 | 9.5± 0.43 |
| 164284 | B4IV | 0.23 | 47.89 | 15820 ± 192 | 22211±573 | 3.89 ± 0.048 | 4.207±0.055 | 240±4 | 276±7 | 5.0± 0.15 |
| 166012 | B9V | 0.48 | 43.4 | 10583 ± 245 | --- | 3.62 ± 0.101 | --- | ---- | ---- | 2.5± 0.07 |
| 168199 | B7V | 0.35 | 47.89 | 12506 ± 303 | 14660±104 | 3.91 ± 0.05 | 3.76±0.019 | --- | 186±8 | 3.5± 0.14 |
| 168270 | B9V | 0.47 | 50.14 | 10490 ± 227 | 10245±34 | 3.79 ± 0.115 | 3.419±0.018 | 30±1.1 | 74±10 | 3.5± 0.08 |
| 169578 | B9V | 0.47 | 54.64 | 10828 ± 48 | 10901±36 | 3.89 ± 0.028 | 3.498±0.014 | 230±2 | 252±9 | 3.0± 0.02 |
| 171301 | B8V | 0.44 | 56.89 | 1250 ± 230 | 12170±82 | 3.6 ± 0.025 | 3.969±0.025 | 10±0.19 | 59±13 | 3.0± 0.55 |
| 171406 | B7V | 0.35 | 50.14 | 12634 ± 81 | 14216±115 | 3.96 ± 0.030 | 3.881±0.022 | — | 248±10 | 3.5± 0.11 |
| 172958 | B8V | 0.45 | 47.89 | 11023 ± 265 | 10727±69 | 3.81 ± 0.115 | 3.577±0.030 | — | 167±12 | 3.5± 0.10 |
| 173087 | B7V | 0.36 | 52.39 | 12612 ± 478 | 14504±111 | 4.03 ± 0.053 | 3.97±0.025 | 100±4 | 91±10 | 3.5± 0.13 |
| 173936 | B7V | 0.36 | 47.89 | 12456 ± 143 | 13489±88 | 3.91 ± 0.026 | 3.989±0.015 | 120±1.5 | 116±8 | 4.0± 0 |
| 174959 | B8IV | 0.41 | 61.39 | 12074 ± 303 | 13499±80 | 4.04 | 3.795±0.012 | 30±0.75 | 52±11 | 3.5± 0 |
| 175426 | B4IV | 0.24 | 41.16 | 15459 ± 55 | 16137±197 | 3.68 | 3.764±0.032 | — | 86±10 | 6.0± 0 |
| 176301 | B8V | 0.41 | 52.39 | 11545 ± 70 | ---- | 3.97± 0.03 | --- | — | --- | 3.0± 0.018 |
| 176318 | B8V | 0.4 | 59.14 | 11962 ± 114 | 13058±67 | 4.03 ± 0.004 | 3.888±0.015 | 120±1.14 | 122±8 | 3.5± 0.04 |
| 177756 | B9V | 0.46 | 61.39 | 10926 ± 74 | 11084±41 | 4.04 ± 0.025 | 3.822±0.016 | 180±1.65 | 170±5 | 3.0± 0.025 |
| 177817 | B9V | 0.44 | 50.14 | 11068 ± 68 | 12387±55 | 3.85 ± 0.02 | 3.64±0.019 | --- | 162±12 | 3.0± 0.04 |
| 178125 | A0IV | 0.53 | 59.14 | 9839 ± 191 | 13120±100 | 3.77 ± 0.044 | 4.078±0.017 | 60±1.35 | 74±7 | 3.0± 0.06 |
| 178329 | B4I | 0.24 | 43.4 | 15360 ± 329 | 15317±208 | 3.76 ± 0.037 | 3.827±0.033 | 10±0.23 | 0±19 | 5.5± 0.15 |
| 179761 | A0IV | 0.54 | 45.65 | 9724 ± 120 | 12746±103 | 3.45 ± 0.055 | 3.469±0.027 | — | 12±6 | 3.5± 0.05 |
| 180163 | B3IV | 0.19 | 34.44 | 17387 ± 238 | 15250±164 | 3.47 ± 0.016 | 3.196±0.026 | — | 37±7 | 8.0± 0.11 |
| 180554 | B7V | 0.41 | 61.39 | 11893 ± 137 | ---- | 4.04 ± 0.005 | --- | 100±1.15 | --- | 3.5± 0.08 |
| 183144 | B6IV | 0.3 | 41.16 | 13534 ± 231 | 14361±126 | 3.70 ± 0.070 | 3.484±0.028 | 230±5.8 | 211±8 | 5.0± 0.10 |
| 184930 | B6IV | 0.29 | 38.92 | 13821 ± 221 | 13148±89 | 3.63 ± 0.032 | 3.621±0.016 | 40±0.7 | 50±9 | 5.0± 0.10 |
| 185423 | B3III | 0.19 | 27.73 | 17133 ± 441 | 16603±328 | 3.24 ± 0.030 | 3.209±0.049 | — | 103±14 | 9.0± 0 |
| 187811 | B4V | 0.24 | 45.65 | 15428 ± 300 | 21331±640 | 3.84 | 4.173±0.062 | 210±4.0 | 242±10 | 5.0± 0.11 |
| 187961 | B7IV | 0.37 | 32.2 | 12170 ± 93 | 16646±441 | 3.46 ± 0.035 | 3.554±0.063 | — | 258±10 | 4.5± 0.05 |
| 188260 | A1IV | 0.55 | 50.14 | 9684 ± 102 | 10363±50 | 3.54 ± 0.032 | 3.592±0.025 | 10±0.13 | 59±8 | 3.0± 0.04 |
| 189944 | B7V | 0.36 | 45.65 | 12406 ± 198 | 14134±175 | 3.85 ± 0.028 | 3.758±0.035 | 30±0.52 | 12±15 | 4.0± 0.07 |
| 190993 | B4V | 0.24 | 45.65 | 15396 ± 335 | ---- | 3.84 ± 0.027 | --- | — | --- | 5.0± 0.12 |
| 191610 | B4V | 0.23 | 43.4 | 15648 ± 246 | ---- | 3.76 ± 0.037 | --- | — | --- | 5.5± 0 |
| 191639 | B2V | 0.13 | 50.14 | 22018 | 29047±1343 | 3.97 | 3.777±0.157 | --- | 152±15 | 9.0± 0 |
| 192276 | B9V | 0.44 | 59.14 | 11112 ± 203 | 13272±155 | 4.03 | 4.088±0.031 | 10±4 | 29±12 | 3.0± 0.06 |
| 192685 | B4V | 0.24 | 41.16 | 15324 ± 248 | 17062±242 | 3.68 ± 0.037 | 3.746±0.033 | 210±0 | 162±11 | 5.5± 0.10 |
| 193432 | B9V | 0.46 | 61.39 | 10957 ± 160 | 10208±53 | 4.04 ± 0.035 | 3.814±0.028 | — | 27±18 | 3.0± 0.05 |
| 193536 | B4V | 0.24 | 38.92 | 15200 ± 323 | ---- | 3.61 ± 0.032 | --- | — | --- | 6.0± 0.14 |
| 195810 | B6IV | 0.3 | 41.16 | 13549 ± 266 | 13146±121 | 3.70 ± 0.037 | 3.646±0.025 | 10±0.22 | 47±10 | 5.0± 0.10 |
| 196504 | B9V | 0.46 | 61.39 | 10861 ± 130 | 10693±59 | 4.04 ± 0.020 | 3.781±0.026 | — | 315±13 | 3.0± 0.04 |
| 196740 | B7V | 0.4 | 61.39 | 11924 ± 600 | 14129±154 | 4.04 ± 0.030 | 3.673±0.030 | — | 276±7 | 3.5± 0.18 |
| 196867 | B9V | 0.46 | 56.89 | 10789 ± 53 | 10568±44 | 3.94 ± 0.032 | 3.572±0.017 | 130±1.2 | 138±5 | 3.0± 0.02 |
| 198183 | B7V | 0.36 | 41.16 | 12319 ± 102 | 14187±137 | 3.71 ± 0.004 | 3.765±0.027 | 140±1.16 | 120±10 | 4.0± 0.06 |
| 199081 | B7V | 0.36 | 47.89 | 12397 ± 190 | ---- | 3.91 ± 0.051 | --- | 30±0.6 | --- | 3.5± 0.05 |
| 205021 | B2V | 0.19 | 54.64 | 20207 ± 1192 | 27784±768 | 3.99 ± 0.025 | 4.261±0.064 | — | 35±5 | 7.0± 0 |
| 206165 | B4III | 0.21 | 27.73 | 16250 | 19887±394 | 3.26 | 2.73±0.046 | — | 58±12 | 8.0± 0 |
| 207516 | B8V | 0.44 | 56.89 | 11089 ± 87 | 12187±80 | 3.99 ± 0.012 | 4.02±0.024 | 80±0.67 | 91±10 | 3.0± 0.03 |
| 208947 | B4V | 0.24 | 50.14 | 15391 ± 389 | ---- | 3.95 ± 0.028 | --- | — | --- | 5.0± 0 |
| 209409 | B8V | 0.4 | 59.14 | 11962 ± 90 | 18389±524 | 4.03 | 4.178±0.065 | 200±1.5 | 224±8 | 3.5± 0 |
| 209819 | B9V | 0.46 | 63.65 | 10922 ± 143 | 12026±45 | 4.04 | 4.161±0.013 | 140±1.8 | 147±8 | 3.0± 0.04 |
| 214923 | B9V | 0.45 | 47.89 | 11000 ± 16 | 11927±89 | 3.80 ± 0.005 | 3.858±0.030 | 210±0.4 | 153±3 | 3.5± 0.03 |
| 217675 | B6IV | 0.3 | 32.2 | 13295 ± 254 | 14458±210 | 3.42 ± 0.033 | 3.195±0.040 | — | 235±11 | 5.0± 0.17 |
| 220575 | A0IV | 0.55 | 41.16 | 9714 ± 182 | 12419±125 | 3.37 ± 0.094 | 3.514±0.034 | — | 18±5 | 3.5± 0.07 |
| 222439 | B9V | 0.46 | 63.65 | 10909 ± 74 | 10632±41 | 4.04 ± 0.004 | 3.875±0.019 | 170±1.16 | 169±4 | 3.0± 0 |
| 224801 | B9V | 0.47 | 65.91 | 10685 ± 120 | ---- | 4.05 | --- | — | --- | 3.0± 0 |
| 224926 | B7V | 0.4 | 63.65 | 12201 ± 123 | 14047±118 | 4.04 | 3.842±0.023 | — | 97±26 | 3.5± 0.04 |
| 225132 | B9V | 0.48 | 52.39 | 10622 ± 60 | 10839±48 | 3.83 ± 0.014 | 3.767±0.014 | 230±1.5 | 249±10 | 3.0±0 |

*HG=Huang and Gies (2008)

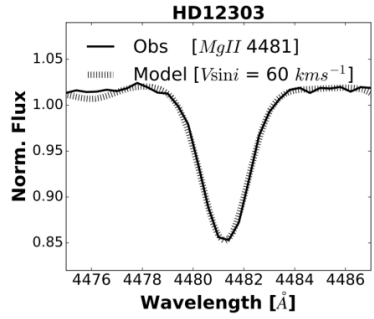
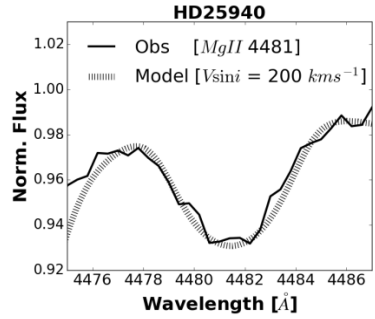
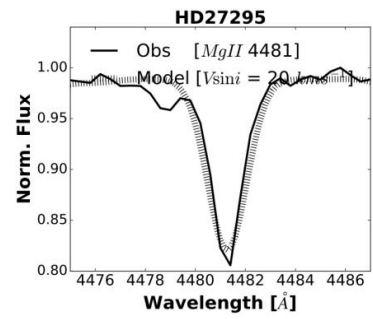
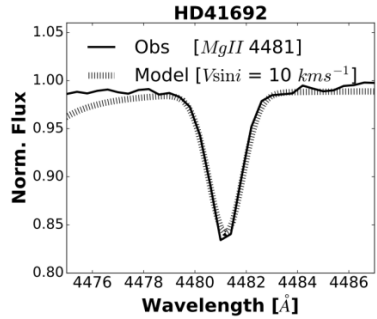
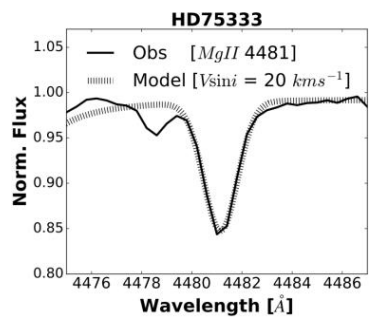
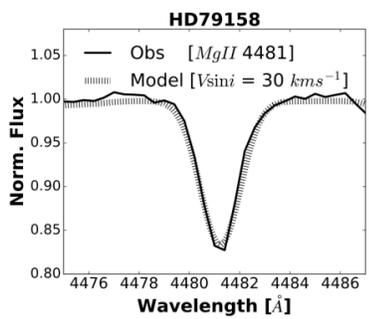
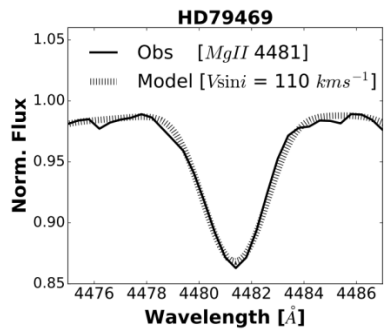
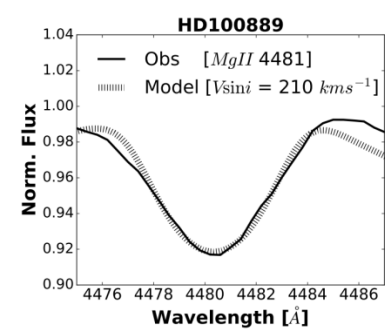
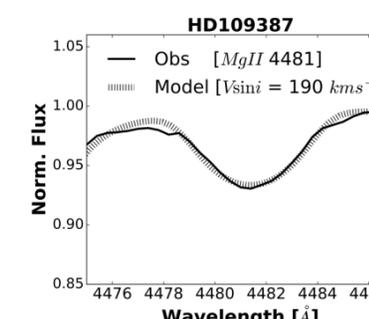
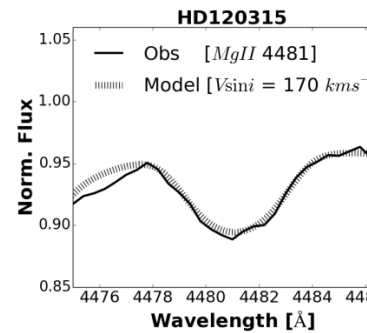
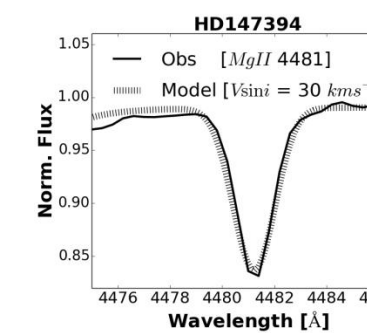
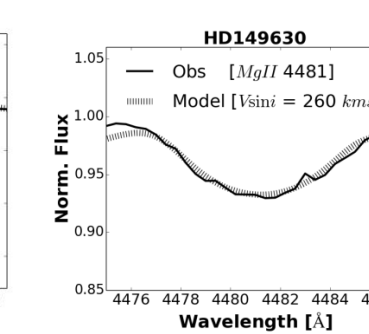
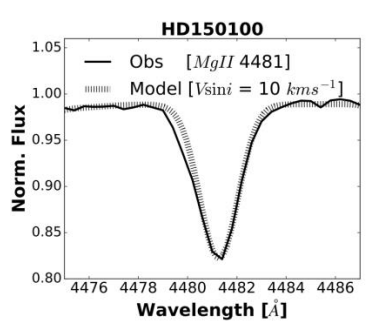
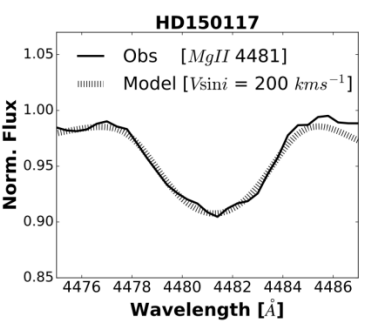
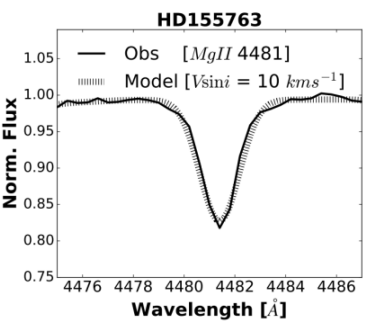

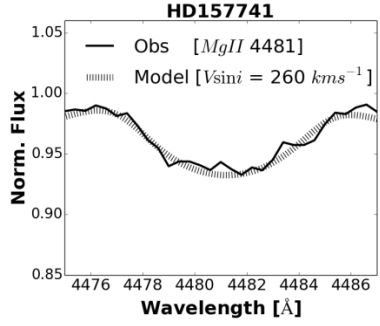
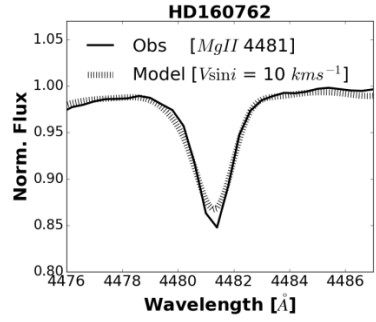
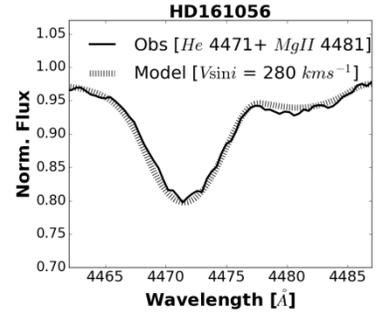
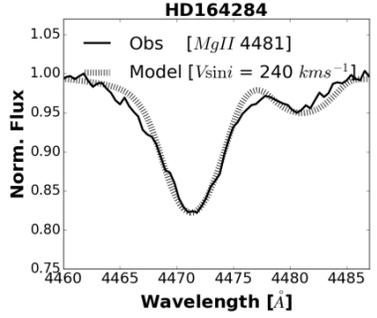
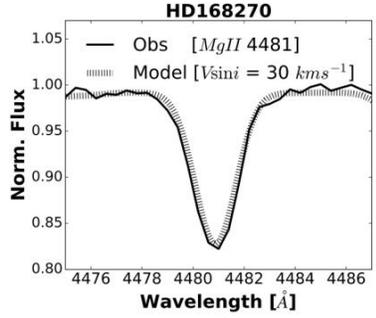
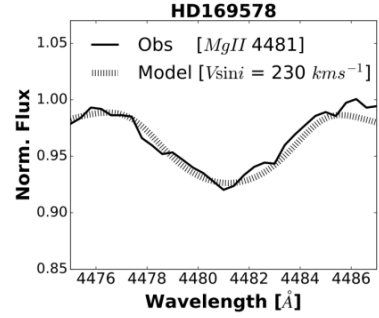
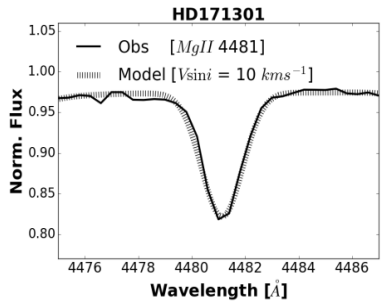
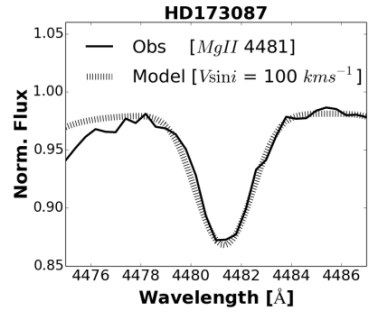
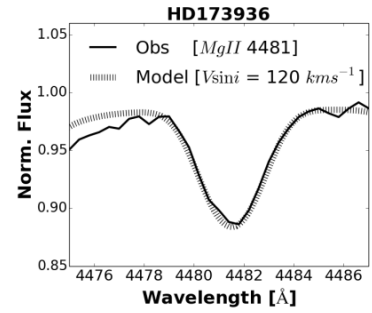
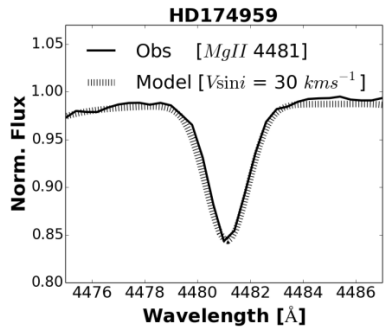
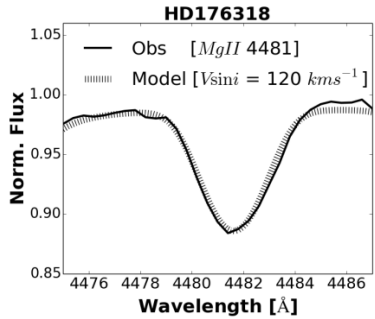
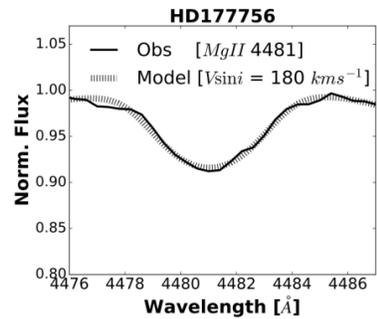
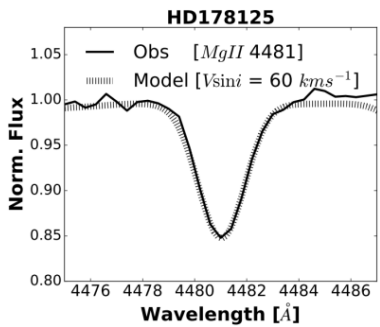
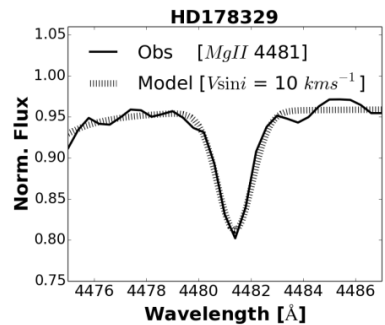
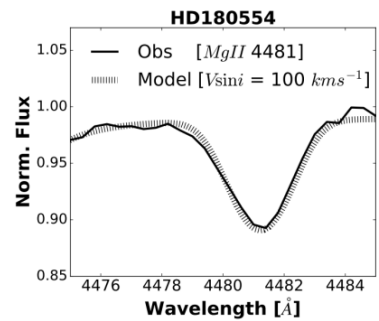

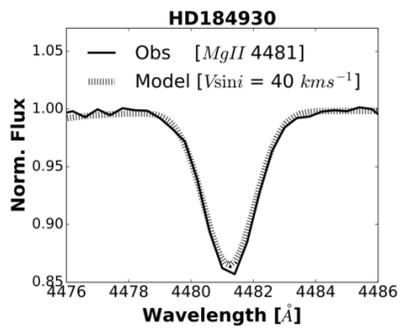
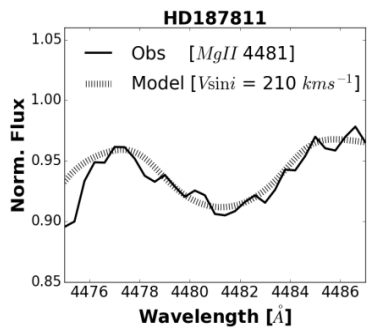
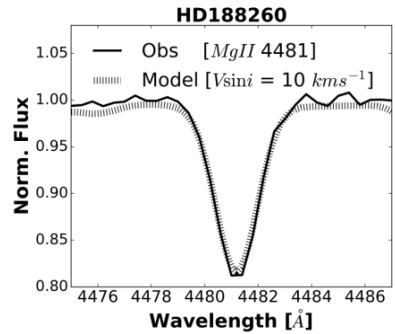
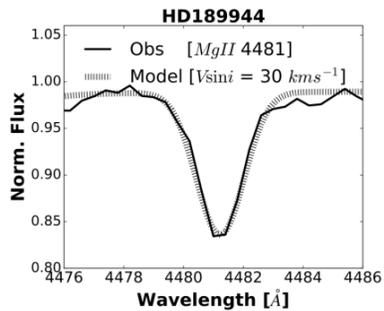
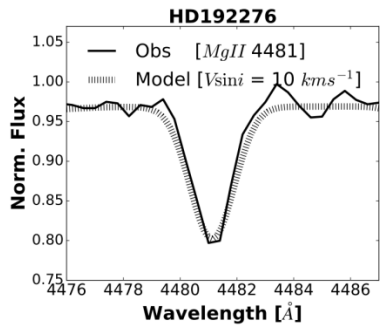
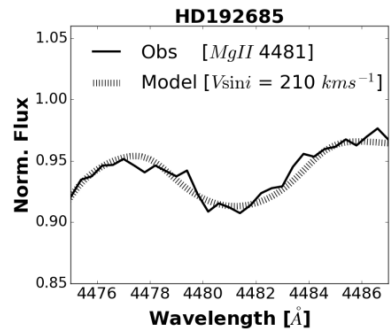
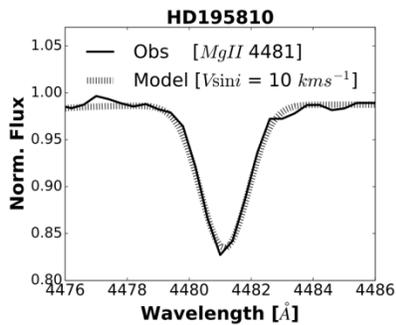
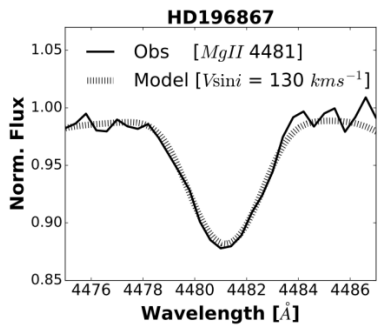
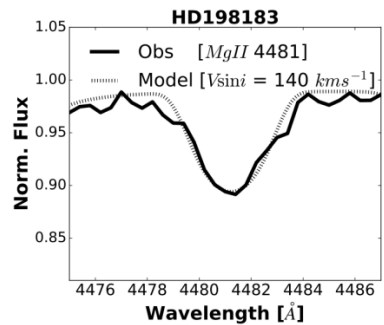
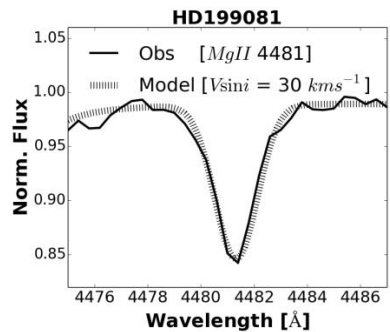
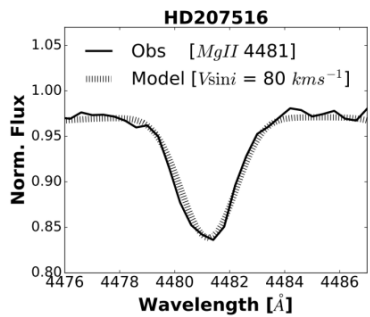
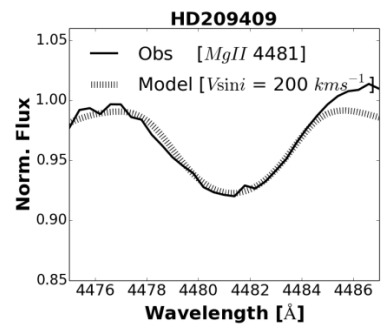

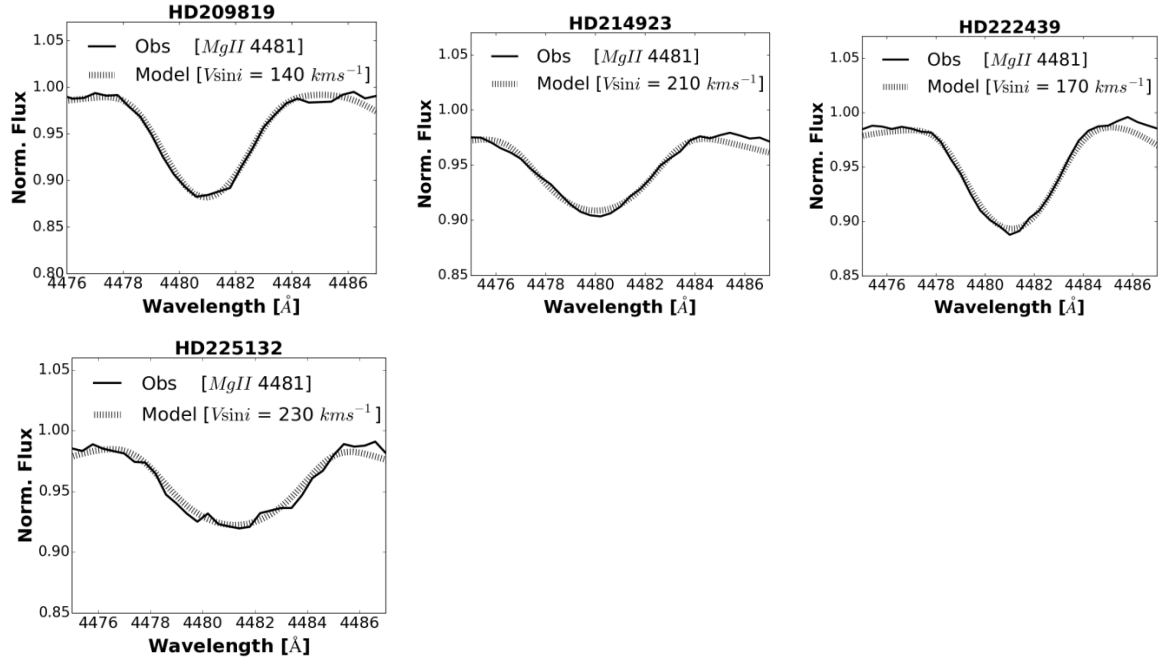

Figure 4: Example of the Mg II 4481 Å line profile fits of some of the program stars. The solid lines are for the observed spectra while the dashed lines are for the LTE/NLTE synthetic ones.

## 4.2 Comparison with Huang & Gies (2008)

Figures (5-7) represent the comparison between the present results with that of Huang and Gies (2008) for the effective temperatures, surface gravities, and rotational velocities. Figures (8) compares the present rotational velocities with those computed by Abt et al. (2003). The great discrepancies between our $T_{eff}$ and log g and those of Huang & Gies (2008) may be attributed to the difference in the methods of analysis. The BCD method can deal efficiently with the low resolution spectra (the NOAO spectral resolution is about FWHM=1.5 $A^0$), where the parameters (λ, D) are derived from the stellar continuum energy distribution by direct measurement. This means on average that these photospheric layers are important to the physical properties that are deeper than those represented by spectral lines (Zorec et al., 2009). The comparison for the rotational velocities gives good agreement as appeared in Figure (7).

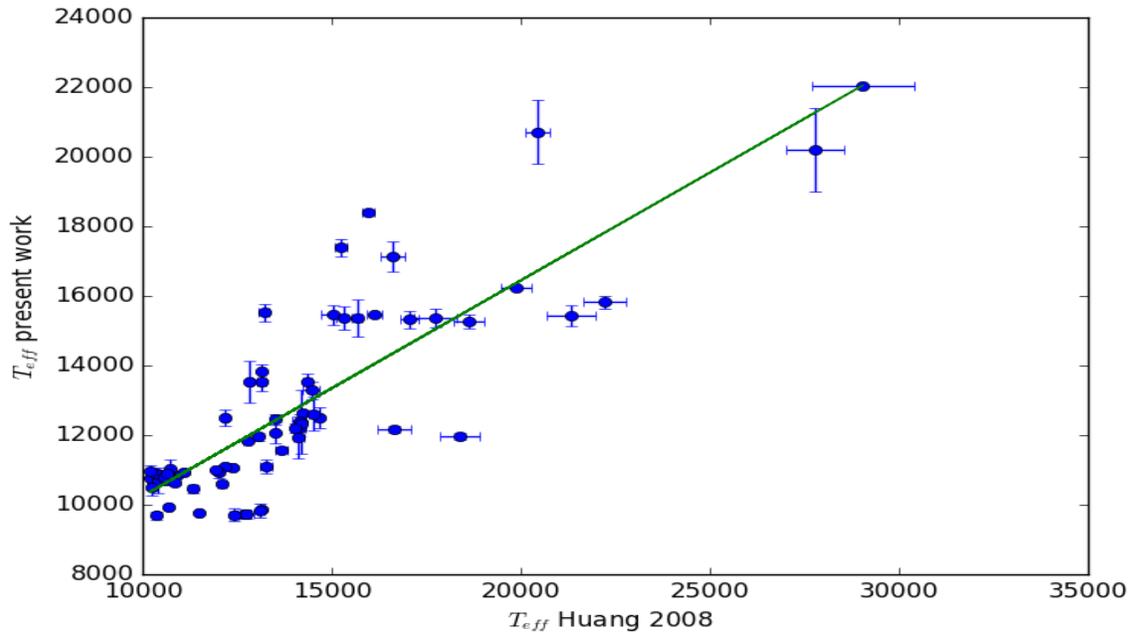

Figure 5: Comparison between our measured log g and those from Huang & Gies (2008).

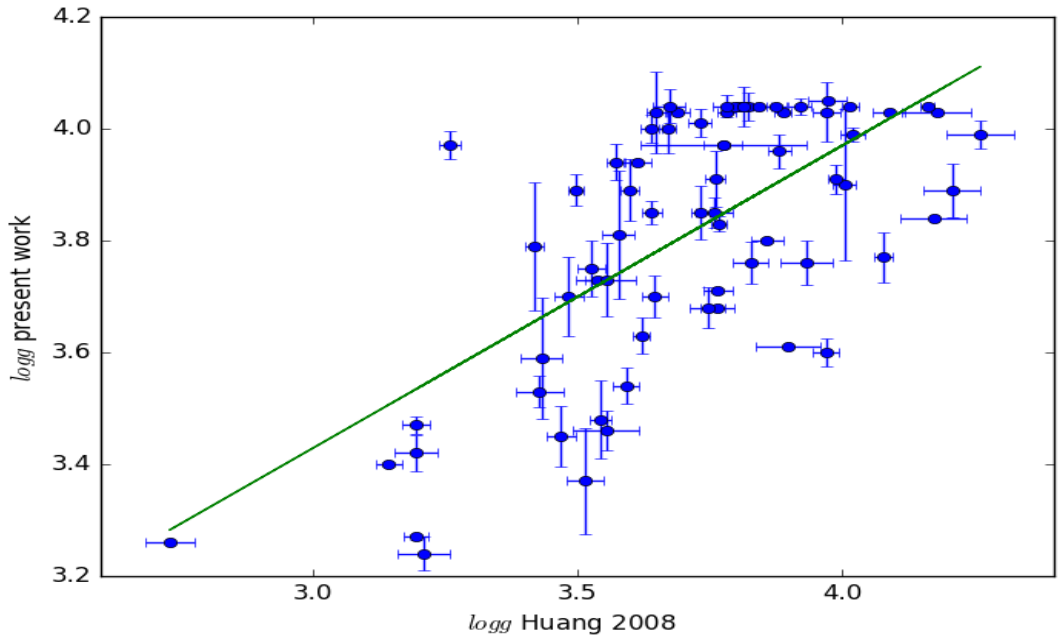

Figure 6: Comparison between our measured Teff and those from Huang & Gies (2008).

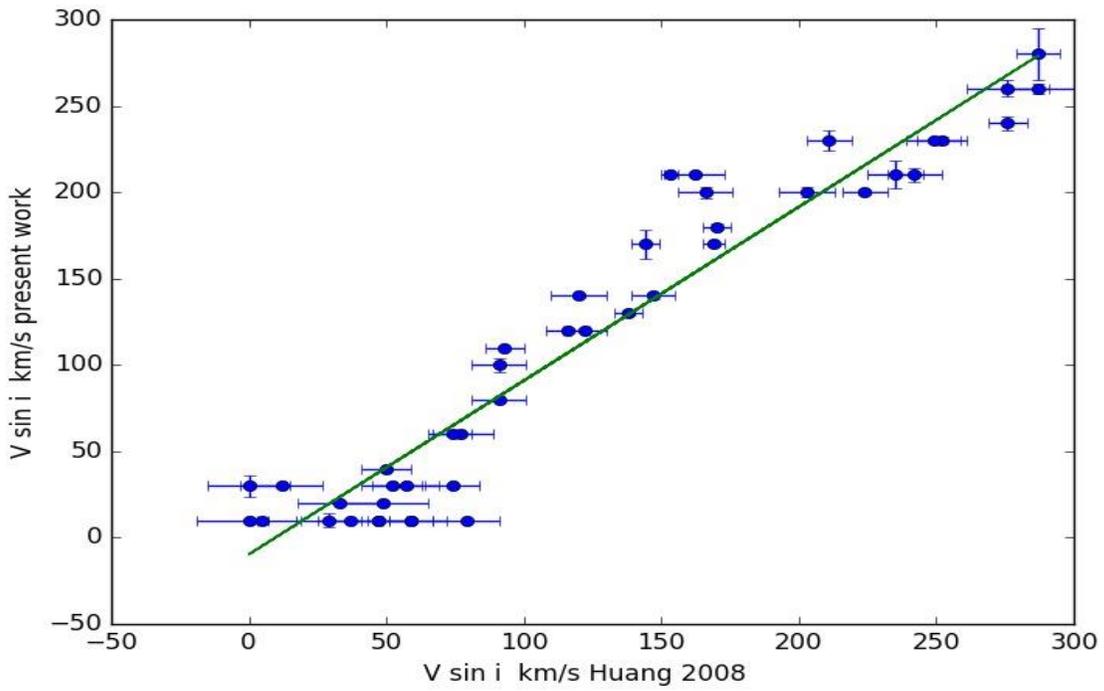

Figure 7: Comparison between our measured v sin i and those from Huang & Gies (2008).

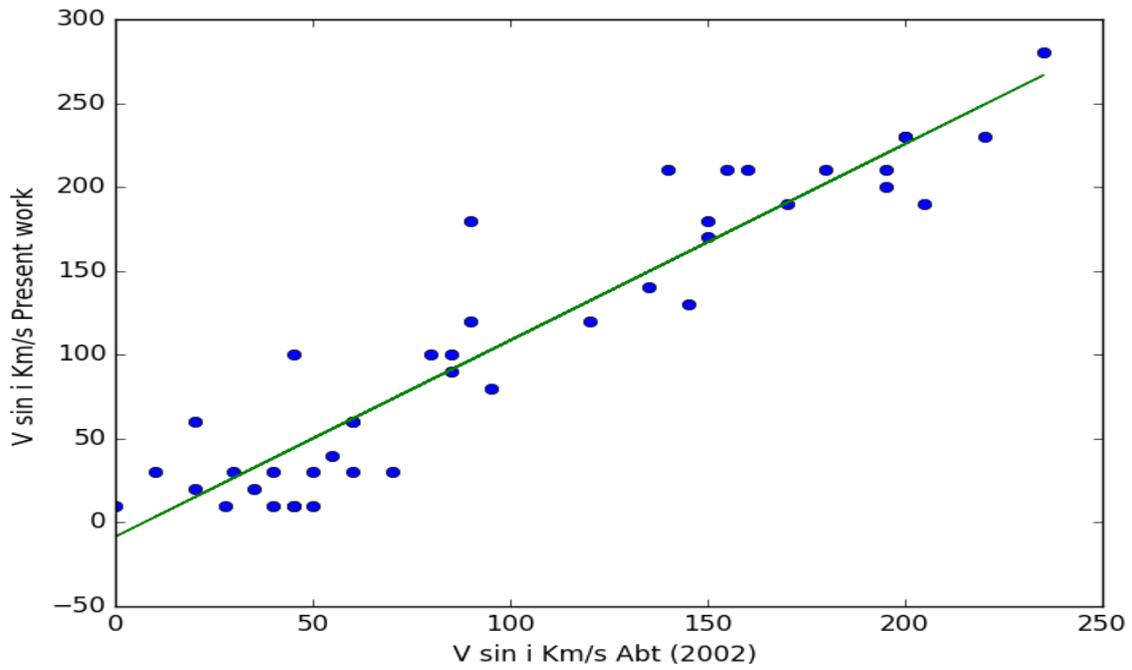

Figure 8: Comparison between our measured v sin i and those from Abt et al. (2002).

## 4.3 Evolution tracks and Stellar masses

Given the stellar parameters, we plotted the effective temperature versus surface gravity along with the evolutionary tracks for the mass range 2.5 - 9 M$_\odot$ of Girardi et al. (2000) with solar metallicity. To estimate the masses and the evolutionary status we located the program stars over the HR diagram.

The errors in the masses are determined by the relation

$$\Delta M = \sqrt{(\Delta M_{T_{eff}})^2 + (\Delta M_{\log g})^2} ,$$

where $\Delta M_{T_{eff}}$ and $\Delta M_{\log g}$ are the changes in the mass due to change in the effective temperature and surface gravity, respectively.

Figure (9) presents the location of studied stars over the HR diagram, while the estimated masses of the stars have been listed in Table 1 (column 8).

Figure (10) shows good agreement of the program stars with the empirical mass-effective temperature relation for main sequence stars by Eker et al. (2018).

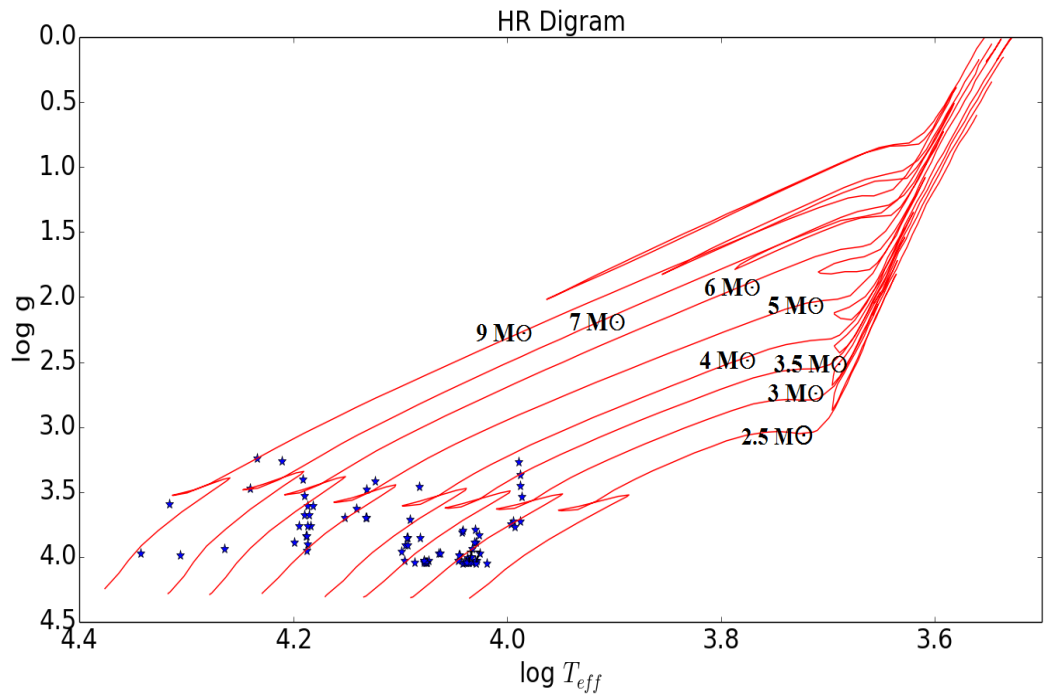

Figure 9: The position of the program stars on the theoretical H-R diagram, based on the stellar evolution models of Girardi et al. (2000).

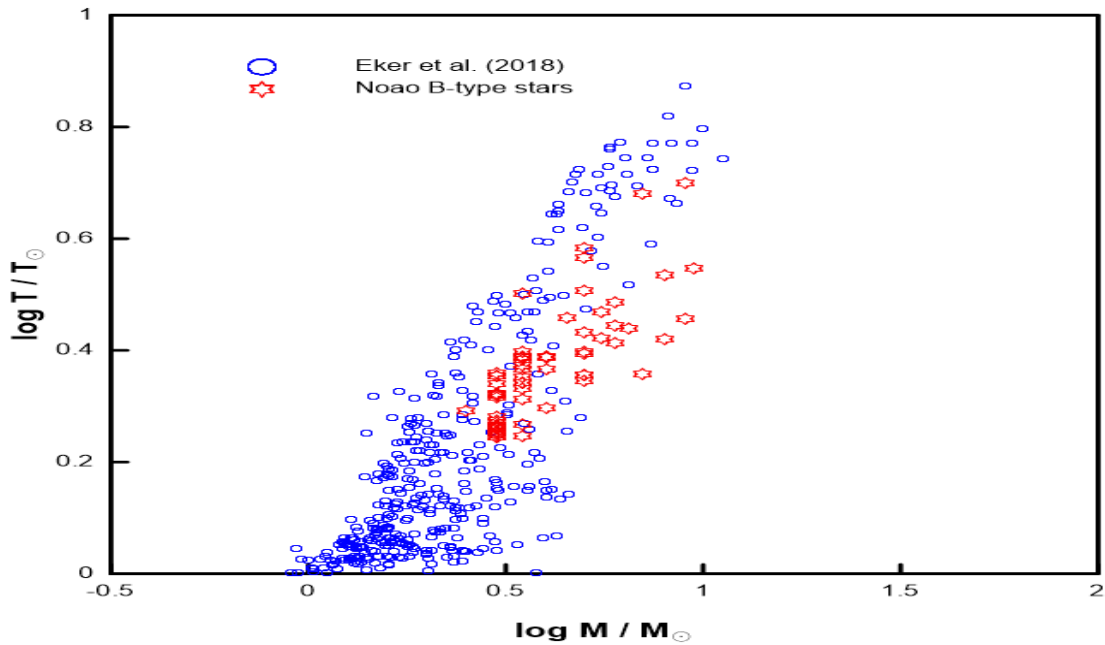

Figure 10: Locations of the program stars on Eker et al. (2018) empirical mass-effective temperature relation for main sequence stars.

## 4.4 Notes on some individual objects

In the following we discuss the previous results for some stars:

**HD4727** is a spectroscopic binary star, classified as B5 spectral type by Lesh (1968). Ducati et al (2011) determined the initial mass of the system as 5.9M, Zorec 2009 determined Teff as 16290 K.

**HD17081** The star was classified as B7 IV by Strom et al (2005), who determined $V_{rot}$ sin i =25 km/s, Prugniel et al (2011) determined Teff as 12722 K and log g =4.2 dex, while Zorec et al 2009 determined the Teff as 14150 K, Fekel& Francis (2003) determined $V_{rot}$ sin i as 21 km/s. Adelman et al (2002) determined Teff as 13100 K and log g as 3.85 dex.

**HD25940** is a B emission star, Abt et al (2002) classified the star as B3Ve and determined $V_{rot}$ sin i as 155 km/s, while Wegner et al (1993) determined $V_{rot}$ sin i as 217 km/s, Slettebak et al (1992) classified stars as B4ve and determined $V_{rot}$ sin i as 200 km/s, Theodossiou&Danezis (1991) determined Teff as 17500 K, while Underhill et al (1979) determined it as 16297 K, Briot (1986) classified the star as B4VIe and determined $V_{rot}$ sin i = 230 Km/s. Vardya (1985) classified the star as B3ve and determined $V_{rot}$ sin i as 230 km/s and mass and radius as 11$M_\odot$ and 6$R_\odot$, respectively.

**HD27295** the star was classified as B9V by Abt et al (1995) who determined the $V_{rot}$ sin i as 10 km/s, Zorec& Royer (2012) classified star as B9IV and determined Teff as 11200 K and $V_{rot}$ sin i as 18 km/s, while McDonald et al (2012) determined Teff as 10438 K, Prugniel et al (2011) determined Teff = 11034 K and log g = 3.99 dex, Behr (2003) determined Teff = 11956 K , log g = 3.92 dex and $V_{rot}$ sin i = 4.0 km/s.

**HD34797** the star was classified as B7V by Babu &Rautela (1978), McDonald et al (2012) determined Teff as 9247 K.

**HD35497** the star was classified as B8 Cr Mn by Stigler et al (2014), while Chen et al (2017) classified the star as B7III. Zorec et al (2009) determined the $T_{eff}$ by using the BCD method as 14910 K and classified the star as B6 IV.

**HD38899** The star was studied by Smith et al (1993) who determined $T_{eff}$ = 10850 K, log g = 4.1 dex with spectral type B9 I, and determined $V_{rot}$ sin i as 30 km/s, while Liu et al (2008) determined Teff as 10903 K and log g 4.0 dex.

**HD41692** the star was classified as B5 IV by Wu et al ( 2011) who determined the Teff 16566 and log g 3.64 dex.

**HD58343** the star was classified as Be star where it was recently studied by Cochetti et al (2020), who determined Teff as 19261 K, log g 3.196 dex, and $V_{rot}$ sin i as 49 km/s, while Arcos (2018) determined Teff = 20000 K, log g =3.56 and $V_{rot}$ sin i =10 km/s, Catanzaro (2013) determined Teff as 18600 K, Frémat (2005) determined Teff = 16389 K and log g = 3.6 dex

**HD74280** ( η Hya ) the star was classified star as B3V by Wu (2011), who determined Teff as 17816 K, log g 3.73 dex. while Schiavon (2007) determined Teff as 18330 K, log g 3.94 dex.

**HD75333** the star was classified as B9mnp by Wu et al (2011), who determined Teff as 11722 K and log g as 4.08 dex.

**HD79469** is a spectroscopic binary star classified as B9.5 by Wu (2011), who determined Teff as 10099 K and log g as 3.8 dex.

**HD87344** the star was classified as B8 by Wu et al (2011), who determined Teff as 11061 and log g as 4.0 dex, while, McDonald et al (2012) determined Teff as 9984 K, Soubiran et al. (2010) determined Teff as 10691 K and log g 3.65 dex.

**HD100889** is a High proper-motion Star classified as B9.5Vn by Wu (2011), who determined Teff as 10563 K and log g as 3.83 dex, while McDonald (2012) determined Teff as 10460 K.

**HD109387** (κ Dra) is a well known Be star, the star classified as B6IIIpe, Soubiran et al. (2010) determined Teff as 13900 K and log g 3.1 dex, Catanzaro (2013) determined $V_{rot}$ sin i as 170 km/s, The binary nature of Kpa Dra has been reported a long time ago by Hill (1926) and later by Miczaika (1950) based on radial velocity (RV) variations. The first orbital solution was obtained by Juza et al. (1991).

**HD120315** (η UMa) is a High proper-motion Star, the star was classified as B3 V by Baines et al (2018), who determined Teff as 15540 K, while Nouh & Saad (2011) determined Teff as 1600 K, log g as 4.0 dex and $V_{rot} \sin i$ = 150 km/s.

**HD147394** (τ Her) is a Pulsating variable Star, Gordon et al (2019) derived Teff as 15400 K, log g =3.86 dex and $V_{rot} \sin i$ = 33 km/s, while McDonald et al 2012 determined Teff as 11372 K.

**HD160762** (ι Her) is a Variable Star of beta Cep type, Nieva et al (2014) classified the star as B3 IV, and determined Teff as 17500 K, log g 3.8 and mass = 6.6 $M_\odot$.

**HD164284** (66 Oph) is a Be Star, Wu et al (2011) classified the star as B2ve, and derived stellar parameters of the star as Teff = 19706 K and log g 3.69 dex.

**HD171406** is a Be star classified as B4 Ve by Guetter (1968), McDonald et al 2012 determined Teff as 11792 K, while Wu et al (2011) derived Teff as 15278 and log g as 3.7 dex.

**HD173087** is a double star system, classified as B5 V by Guetter (1968), McDonald et al (2012) determined Teff as 12824, while Wu et al (2011) derived Teff as 16182 K and log g as 3.76 dex.

**HD178125** (18 Aql) is Eclipsing binary star, classified as B8III by Osawa (1959), McDonald et al (2012) determined Teff as 10588 K, while Wu et al (2011) derived Teff as 12287 K and log g as 4.18 dex.

**HD187811** (12 Vul) is a Be Star, the star was classified as B2.5 Ve by Wu et al (2011), who derived Teff as 18859 K and log g as 3.36 dex.

**HD 191639** is a Be Star, the star was classified as B1 V by Wu et al (2011), who derived Teff as 20368 and log g as 2.82 dex.

**HD205021** (β Cep) is a Variable Star of beta Cep type, the star was classified as B2III by Wu et al (2011), who derived Teff as 24636 and log g as 3.64 dex.

**HD209409** (o Aqr) is a Be Star, the star was classified as B7IVe by Wu et al (2011), who derived Teff as 12221 K and log g as 2.62 dex, while Cochetti et al (2020) classified the star as B6IV-V and derived stellar parameters of the star as, Teff = 14191, log g = 2.873 dex and $V_{rot} \sin i$ = 240 km/s.

**HD220575** the star was studied by Prugniel et al (2011), who determined Teff as 12241 K and log g as 4.09 dex.

## 5. Discussion and Conclusion

In the present work, we determined the fundamental parameters for 83 Filed B-stars by adopting the modified BCD method, which gives direct estimation to $T_{eff}$ and log g, This method is well recommended for low resolution spectra as in the case of NOAO INDO US library. Based on the obtained $T_{eff}$ and log g we calculated the v sin i by fitting the observed MgII 4481 Å line with theoretical models. Comparing our results with Huang & Gies (2008), we could see that our $T_{eff}$ values are below their value because they are based on a model atmosphere that ignores line blanketing, which results in effective temperatures that can be slightly hotter than the actual values described above (Huang & Gies, 2006). For the BCD method, direct measurement of the stellar continuum distribution energy produces the parameters (λ, D). This means on average that these photospheric layers are important to the physical properties that are deeper than those represented by spectral lines.

## Acknowledgment:

This study was made by using the NOAO IndoU.S. Library of Coude ´ Feed Stellar Spectra archive. This material is based on work supported by the National Science Foundation under grant AST-0606861. The research is supported by the Academy of Science Research and technology (ASRT) under the project titled "Investigating the Spectral and Photometric Behavior of the Cataclysmic Variables in Multi- Wavelength". The research is partially supported by the Science and Technology Development Fund STDF No. 5217.

## References

Abt, H. A., Levato, H., Grosso, M., Rotational Velocities of B Stars, APJ. 573 (1) (2002) 359–365. doi:10.1086/340590.

Abt, H. A., Morrell, N. I., The Relation between Rotational Velocities and Spectral Peculiarities among A-Type Stars, APJS 99 (1995) 135. doi:10.1086/192182.

Adelman, S. J., Pintado, O. I., Nieva, M. F., Rayle, K. E., J. Sanders, S. E., On the effective temperatures and surface gravities of superficially normal main sequence band B and A stars, AAP 392 (2002) 1031–1037. doi:10.1051/0004-6361:20020889.

Arcos, C., Kanaan, S., Chávez, J., Vanzi, L., Araya, I., Curé, M., Stellar parameters and H α line profile variability of Be stars in the BeSOS survey, MNRAS 474 (4) (2018) 5287–5299. arXiv:1711.08675, doi:10.1093/mnras/stx3075.

Athay, R. G., Radiation transport in spectral lines., Vol. 1, Springer Netherlands, 1972.

Barbier, D., Chalonge, D., Etude du rayonnement continu de quelques ´etoiles entre 3 100 et 4 600 A (4e Partie-discussion générale)., Annales d'Astrophysique 4 (1941) 30.

Baines, E. K., Armstrong, J. T., Schmitt, H. R., Zavala, R. T., Benson, J. A., Hutter, D. J., Tycner, C., van Belle, G. T., VizieR Online Data Catalog: Fundamental parameters of 87 stars from the NPOI (Baines+, 2018), VizieR Online Data Catalog (2018) J/AJ/155/30.

Babu, G. S. D., Rautela, B. S., Effective Temperatures, Radii and Bolometric Magnitudes of Ap and Am Stars, APSS 58 (1) (1978) 245–254. doi:10.1007/BF00645389.

Behr, B. B., Rotation Velocities of Red and Blue Field Horizontal-Branch Stars, APJS 149 (1) (2003) 101–121. arXiv:astro-ph/0307232, doi:10. 1086/378352.

Briot, D., Rotational velocity of Be stars correlated with emission characteristics., AAP 163 (1986) 67–76.

Cowley, A., Spectral classification of the bright B8 stars., AJ 77 (1972) 750–755. doi:10.1086/111348.

Chalonge, D., Divan, L., Recherches sur les spectres continus stellaires. V. Etude du spectre continu de 150 etoiles entre 3150 et 4600 A., Annales d'Astrophysique 15 (1952) 201.

Ducati, J. R., Penteado, E. M., Turcati, R., The mass ratio and initial mass functions in spectroscopic binaries, AAP 525 (2011) A26. doi:10.1051/ 0004-6361/200913895.

Catanzaro, G., Spectroscopic atlas of Hα and Hβ in a sample of northern Be stars, AAP 550 (2013) A79. arXiv:1212.6608, doi:10.1051/0004-6361/ 201220357.

Chen, P. S., Liu, J. Y., Shan, H. G., A New Photometric Study of Ap and Am Stars in the Infrared, AJ 153 (5) (2017) 218. doi:10.3847/1538-3881/ aa679a.

Cochetti, Y. R., Zorec, J., Cidale, L. S., Arias, M. L., Aidelman, Y, Torres, A. F, Y. ,Frémat, A. Granada, Be and Bn stars: Balmer discontinuity and stellar-class relationship, AAP 634 (2020) A18. arXiv:1912.12994, doi:10.1051/0004-6361/201936444.

Eker, Z.; Bakış, V.; Bilir, S.; Soydugan, F.; Steer, I.; Soydugan, E.; Bakış, H.; Aliçavuş, F.; Aslan, G.; Alpsoy, M., Interrelated main-sequence mass-luminosity, mass-radius, and mass-effective temperature relations, 2018, MNRAS, 479, 5491.

Falcón-Barroso, J., Sánchez-Blázquez, P., Vazdekis, A., Ricciardelli, E., Cardiel, N., Cenarro, A. J., Gorgas, J., Peletier, R. F., An updated MILES stellar library and stellar population models, AAP 532 (2011) A95. arXiv:1107.2303, doi:10.1051/0004-6361/201116842.

Fekel, F. C., Rotational Velocities of B, A, and Early-F Narrow-lined Stars, PASP 115 (809) (2003) 807–810. doi:10.1086/376393.

Frémat, Y., Zorec, J., Hubert, A. M., Floquet, M., Effects of gravitational darkening on the determination of fundamental parameters in fast-rotating B-type stars, AAP 440 (1) (2005) 305–320. arXiv:astro-ph/0503381, doi:10.1051/0004-6361:20042229.

Gray, R. O., The calibration of Stromgren photometry for A, F and early G supergiants. III. The A and early F supergiants., AAP 265 (1992) 704–710.

Gray, D. F., The observation and analysis of stellar photospheres, Cambridge University Press, 2015.

Girardi, L., Bressan, A., Bertelli, G., Chiosi, C., Evolutionary tracks and isochrones for low- and intermediate-mass stars: From 0.15 to 7 M and from Z=0.0004 to 0.03, AAPS 141 (2000) 371–383. arXiv:astro-ph/ 9910164, doi:10.1051/aas:2000126.

Gordon, K. D., Gies, D. R., Schaefer, G. H., Huber, D., Ireland, M., Angular Sizes, Radii, and Effective Temperatures of B-type Stars from Optical Interferometry with the CHARA Array, APJ 873 (1) (2019) 91. doi:10.3847/1538-4357/ab04b2.

Guetter, H. H., Spectral Classifications of 239 Early-Type Stars, PASP. 80 (473) (1968) 197. doi:10.1086/128611.

Hill, S. N., The orbits of two spectroscopic binaries, Publications of the Dominion Astrophysical Observatory Victoria 3 (1926) 349–363.

Huang, W., Gies, D. R., Stellar Rotation in Field and Cluster B Stars, APJ 683 (2) (2008) 1045–1051. arXiv:0805.2133, doi:10.1086/590106.

Jensen, K. S., Spectral classification in the MK system of 167 northern HD stars., AAPS 45 (1981) 455–458.

Juza, K., Harmanec, P., Hill, G. M., Tarasov, A. E., Matthews, J. M., Tuominen, I., Yang, S., Properties and Nature of Be Stars. 16. Closer to a Solution of the Puzzle of 5 κ Dra?, Bulletin of the Astronomical Institutes of Czechoslo- vakia 42 (1991) 39.

Kahn, F. D., Woltjer, L., Intergalactic Matter and the Galaxy., APJ 130 (1959) 705. doi:10.1086/146762.7

Kurucz, R. L., Synthetic template spectra, Highlights of Astronomy 10 (1995) 407.

Lanz, T., Hubeny, I., A Grid of NLTE Line-blanketed Model Atmospheres of Early B-Type Stars, APJS 169 (1) (2007) 83–104. arXiv:astro-ph/ 0611891, doi:10.1086/511270.

Le Borgne, J. F., Bruzual, G., Pell´o, R., Lan¸con, A., Rocca-Volmerange, B., Sanahuja, B., Schaerer, D., Soubiran, C., V´ılchez-G´omez, R., STELIB: A library of stellar spectra at R ˜2000, AAP 402 (2003) 433–442. arXiv: astro-ph/0302334, doi:10.1051/0004-6361:20030243.

Lesh, J. R., The Kinematics of the Gould Belt: an Expanding Group?, APJS 17 (1968) 371. doi:10.1086/190179.

Liu, G. Q., Deng, L., Ch´avez, M., Bertone, E., Davo, A. H., MataCh´avez, M. D., A

spectroscopic study of the blue stragglers in M67, MNRAS 390 (2) (2008) 665–674. arXiv:0811.2553, doi:10.1111/j.1365-2966. 2008.13741.x. [44] G. R. Miczaika, Uber das System ¨ ψ Persei. Mit 6 Textabbildungen, ZAP. 28 (1950) 43.

McDonald, I., Zijlstra, A. A., Boyer, M. L., Fundamental parameters and infrared excesses of Hipparcos stars, MNRAS 427 (1) (2012) 343–357. arXiv:1208.2037, doi:10.1111/j.1365-2966.2012.21873.

Nieva, M.-F., Przybilla, N., Fundamental properties of nearby single early B-type stars, AAP 566 (2014) A7. arXiv:1412.1418, doi:10.1051/ 0004-6361/201423373.

Prugniel, P., Vauglin, I., Koleva, M., The atmospheric parameters and spectral interpolator for the MILES stars, AAP 531 (2011) A165. arXiv: 1104.4952, doi:10.1051/0004-6361/201116769.

S´anchez-Bl´azquez, P., Peletier, R. F., Jim´enez-Vicente, J., Cardiel, N., Cenarro, A. J., Falc´on-Barroso, J., Gorgas, J., Selam, S., Vazdekis, A., Mediumresolution Isaac Newton Telescope library of empirical spectra, MNRAS 371 (2) (2006) 703–718. arXiv:astro-ph/0607009, doi:10.1111/j.1365-2966.2006.10699.x.

Sokolov, N. A., The determination of T eff of B, A and F main sequence stars from the continuum between 3200 A and 3600 A., AAPS 110 (1995) 553.

Soubiran, C., Le Campion, J. F., Cayrel de Strobel, G., Caillo, A., The PASTEL catalogue of stellar parameters, AAP 515 (2010) A111. arXiv:1004.1069, doi:10.1051/0004-6361/201014247.

Shokry, A., Rivinius, T., Mehner, A., Martayan, C., Hummel, W., Townsend, R. H. D., M´erand, A., Mota, B., Faes, D. M., Hamdy, M. A., Beheary, M. M., Gadallah, K. A. K., Abo-Elazm, M. S., Stellar parameters of Be stars observed with X-shooter, AAP 609 (2018) A108. arXiv:1711.02619, doi:10.1051/0004-6361/201731536.

Slettebak, A., Collins, I., George, W., Truax, R., Physical Properties of Be Star Envelopes from Balmer and Fe II Emission Lines, APJS 81 (1992) 335. doi:10.1086/191696.

Smith, K. C., Elemental abundances in normal late-B and HgMn stars from co-added IUE spectra. II. Magnesium, aluminium and silicon., AAP 276 (1993) 393.

Stigler, C., Maitzen, H. M., Paunzen, E., Netopil, M., Spectrophotometric analysis of the 5200 ˚

A region for peculiar and normal stars, AAP 562 (2014) A65. arXiv:1402.1021, doi:10.1051/0004-6361/201322300.

Schiavon, R. P., Population Synthesis in the Blue. IV. Accurate Model Predictions for Lick Indices and UBV Colors in Single Stellar Populations, APJS 171 (1) (2007) 146–205. arXiv:astro-ph/0611464, doi:10.1086/511753.

Strom, S. E., Wolff, S. C.,. Dror, D. H. A, B Star Rotational Velocities in h and χ Persei: A Probe of Initial Conditions during the Star Formation Epoch?, AJ 129 (2) (2005) 809–828. arXiv:astro-ph/0410337, doi:10.1086/426748.

Saad, S. M., Nouh, M. I., Model Atmosphere Analysis of Some B-Type Stars, International Journal of Astronomy and Astrophysics 1 (2) (2011) 44–50. doi:10.4236/ijaa.2011.12007.

Theodossiou, E., Danezis, E., The Stellar Temperature Scale for Stars of Spectral Types from o8 to f6 and the Standard Deviation of the MK Spectral Classification, APSS 183 (1) (1991) 91–115. doi:10.1007/BF00643019.

Townsend, R. H. D., Owocki, S. P., Howarth, I. D., Be-star rotation: how close to critical?, MNRAS 350 (1) (2004) 189–195. arXiv:astro-ph/ 0312113, doi:10.1111/j.1365-2966.2004.07627.

Underhill, A. B., Divan, L., Prevot-Burnichon, M. L., Doazan, V., Effective temperatures, angular diameters, distances and linear radii for 160 O and B stars., MNRAS 189 (1979) 601–605. doi:10.1093/mnras/189.3.601.

Vazdekis, A., Ricciardelli, E., Cenarro, A. J., Rivero-Gonz´alez, J. G., D´ıaz-Garc´ıa, L. A., Falc´on-Barroso, J., MIUSCAT: extended MILES spectral coverage - I. Stellar population synthesis models, MNRAS 424 (1) (2012) 157–171. arXiv:1205.5496, doi:10.1111/j.1365-2966.2012.21179.x.

Valdes, F., Gupta, R., Rose, J. A., Singh, H. P., Bell, D. J., The Indo-US Library of Coud´e Feed Stellar Spectra, APJS 152 (2) (2004) 251–259. arXiv:astro-ph/0402435, doi:10.1086/386343.

Vardya, M. S., Stellar rotation and mass loss in O and B stars., APJ 299 (1985) 255–264. doi:10.1086/163696.

Wegner, W., Papaj, J., Krelowski, J., Rotational Velocity of Be Stars Correlated with Extinction Law, ACTAA 43 (1993) 53–60.

Wu, Y., Singh, H. P., Prugniel, P., Gupta, R., Koleva, M., Coud´e-feed stellar spectral library – atmospheric parameters, AAP 525 (2011) A71. arXiv: 1009.1491, doi:10.1051/0004-6361/201015014.

Zorec, J., Royer, F., Rotational velocities of A-type stars. IV. Evolution of rotational velocities, AAP 537 (2012) A120. arXiv:1201.2052, doi: 10.1051/0004-6361/201117691.

Zorec, J., Cidale, L., Arias, M. L., Fr´emat, Y., Muratore, M. F., Torres, A. F., Martayan, C., Fundamental parameters of B supergiants from the BCD system. I. Calibration of the (λ 1, D) parameters into Teff, AAP 501 (1) (2009) 297–320. arXiv:0903.5134, doi:10.1051/0004-6361/200811147.